\documentclass[preprint,12pt]{elsarticle}
\usepackage{color,amsthm,url}
\usepackage{amssymb,amsmath}

\newtheorem{remark}{Remark}
\def\im{\mbox{Im\,}}
\def\re{\mbox{Re\,}}
\journal{Applied Numerical Mathematics}

\begin{document}

\begin{frontmatter}



\title{Manakov Model with Gain/Loss Terms and $N$-soliton Interactions: Effects of Periodic Potentials}


\author{V. S. Gerdjikov$^{1,2}$, M. D. Todorov$^3$}

\address{$^1$ Institute of Mathematics and Informatics,\\
 Bulgarian Academy of Sciences, \\
8 Acad. G. Bonchev str.,  1113 Sofia, Bulgaria \\ [5pt]
$^2$Department of Applied Mathematics, \\
National Research Nuclear University MEPHI, \\
31 Kashirskoe Shosse, 115409,  Moscow, Russian Federation \\[5pt]
$^3$Department of Applied Mathematics and Computer Science,\\
Technical University of Sofia, \\
8 Kliment Ohridski, Blvd., 1000 Sofia, Bulgaria \\
}

\begin{abstract}
We analyze the dynamical behavior of the $N$-soliton train in the adiabatic approximation
of the perturbed nonlinear Schr\"odinger equation (NLSE) and the Manakov model.
The perturbations include the simultaneous by a periodic external potential,
and linear and nonlinear gain/loss terms. We derive the corresponding perturbed complex Toda chain
(PCTC) models for both NLS and Manakov model. We show that
the soliton interactions dynamics for the  PCTC models compares favorably to full
numerical results of the original perturbed NLSE and Manakov model.

\end{abstract}


\end{frontmatter}


%\section{}
\label{}


\section{Introduction}
%%%%%%%%%%%%%%%%%%%%%%%%%%%%%%%%%%%%%%%%%%%%%%%%%%%%%%%%

 The NLSE \cite{ZaSh0}
\begin{equation}\label{eq:nls}\begin{split}
i\frac{\partial u}{ \partial t} + \frac{1}{2} \frac{\partial u}{ \partial x^2 } +|u|^2 u(x,t)=0,
\end{split}\end{equation}
and the Manakov model \cite{Man}
\begin{equation}\label{eq:nlsM}\begin{split}
i\frac{\partial \vec{u}}{ \partial t} + \frac{1}{2} \frac{\partial \vec{u}}{ \partial x^2 } +
(\vec{u}^\dag, \vec{u}) \vec{u}(x,t)=0, \qquad \vec{u} = (u_1, u_2)^T
\end{split}\end{equation}
were among the first nonlinear equations, which were shown to possess Lax representation and $N$-soliton solutions.
These results allowed one to explain phenomena taking place in optical media with Kerr nonlinearity.
An important step in this was based on the analysis of soliton interactions. In their very first
paper \cite{ZaSh0} on NLSE Zakharov and Shabat calculated the large time asymptotics of the $N$-soliton solutions
and established that the soliton interactions are purely elastic.

During the following decades revealed a number of new applications of both NLSE and Manakov model in physics.
It is not possible to describe them in detail, so we just list a few references on them. These include:
i) nonlinear optics \cite{UST,ab1,chap01:ablowitz,has2,emergent,siambook,chap01:kiag}; ii) Bose-Einstein condensate
\cite{emergent,siambook,becbook1,becbook2} and in other fields of physics
\cite{dodd,acnewell,Rossi1,Rossi2,nonlinsc,chap01:sulem} and the numerous references therein.

Most of these applications must take into account various perturbative
terms such as external potentials \cite{GT,siambook,becbook1,124a,126a},
effects of cross-channel modulation \cite{GTK2} and others. Such perturbations violate integrability
which prevents the use of exact methods based on the inverse scattering method. The disadvantage
is due to the fact, that  each soliton is parametrized by 4 parameters so
configurations involving two or more  solitons  involve too
many parameters to be studies in detail.

A way out of this difficulty could be based on the adiabatic approach to soliton interactions
was proposed by Karpman and Soloviev \cite{Karp}.  Their idea was to consider the NLSE (\ref{eq:nls})
with initial condition which is sum of two solitons  and then derive a dynamical system for the soliton parameters.
\cite{Karp}. The method involved small parameter $\varepsilon$ -- the `intersection` between the two solitons,
and all calculations neglect terms of order higher than $\varepsilon$.

The method was later generalized to deal with $N$-soliton configurations \cite{PRL,PRE,PLA}
showing that relevant dynamical system describing $N$-soliton interactions ia a Toda chain with complex-valued
dynamical variables -- the complex Toda chain (CTC). Of course this method could
be applied to systems, that are close to integrable ones, like the perturbed  NLSE and Manakov systems, and that satisfy
the adiabatic approximation (see below Section 2). In \cite{VG1,GDM,MCS} the the method of adiabatic approximation was
generalized also for the Manakov system. The result is a modification of the CTC, see Section 3 below.

The present paper can be viewed as a sequel of the papers  \cite{GT,GTK1} co-authored by us, as well as
the most recent one \cite{CGGT} where we analyzed the effects of linear/nonlinear gain/loss terms.
The first difficulty that we encounter on this way is in the fact that gain/loss terms with generic
coefficients typically lead to sharp rise/decay of the soliton amplitudes. In such situation we quickly
come out of the range of application of the adiabatic approximation.

So it will be important to find out constraints on the gain/loss terms coefficients which would be
compatible with the adiabaticity condition. One such constraint was found in \cite{CGGT} for the
$N$-soliton interactions of the scalar NLSE. There we have derived a generalization of the CTC
taking into account gain/loss terms. Here we generalize these results also to the Manakov model,
demonstrating that such perturbations do not affect substantially the evolution of the polarization
vectors. We also note that the method allows to take into account possible $t$-dependence of the
coefficients, which could be used to stabilize their effect.

In Section 2 below we briefly detail the derivation of the Karpman-Soloviev equations and the CTC for the
scalar NLSE. Then we reproduce  the results of effects of gain/loss terms in \cite{CGGT}.
After we have stabilized the effects of gain/loss terms we can also take into account additional perturbations, such as
external potentials.  In Section 3
we generalize these results to the Manakov model and demonstrate that the effects of gain/loss terms ire
basically the same as in the scalar case. We also briefly consider the situations when the gain/loss
coefficients could be time-dependent. In the last Section 4 we compare the numerical results from the
perturbed NLSE and Manakov models to the (also numerica) solutions of the perturbed CTC model. We
do this on the example of 5-soliton trains and find that if the gain/loss terms are compatible with
the adiabatic approximation, then the effects of the periodic external potentials are similar to the ones
studied before \cite{GT,124a,126a}.
In the last Section we have collected some concluding remarks and our views on future activities. The Appendix
contains several types of useful integrals used in the calculations.

\section{Preliminaries}

%%%%%%%%%%%%%%%%%%%%%%%%%%%%%%%%%%%%%%%%%%%%%%%%%%%%%%%%
\subsection{The adiabatic approximation}
%%%%%%%%%%%%%%%%%%%%%%%%%%%%%%%%%%%%%%%%%%%%%%%%%%%%%%%%

Lets us consider an $N$-soliton train as an initial condition to the perturbed NLSE.
By $N$-soliton train we mean a chain of several well-separated solitons whose
parameters comply with the adiabatic approximation:
\begin{equation}\label{eq:Nst}\begin{aligned}
u(x,t=0) &= \sum_{k=1}^N u_k(x,t=0), &\qquad u_k(x,t) &= {2\nu_k e^{i\phi_k}\over \cosh(z_k)}  ,\\
z_k &= 2\nu_k (x-\xi_k(t)), &\qquad \xi_k(t) &=2\mu_k t +\xi_{k,0}, \\
\phi_k &= {\mu_k \over \nu_k} z_k + \delta_k(t), &\qquad \delta_k(t) &=2(\mu_k^2+\nu_k^2) t +\delta_{k,0}.
\end{aligned}\end{equation}
The adiabatic approximation holds true if the soliton parameters satisfy \cite{Karp}:
\begin{eqnarray}\label{eq:ad-ap}
&& |\nu _k-\nu _0| \ll \nu _0, \quad |\mu _k-\mu _0| \ll \mu _0,
\quad |\nu _k-\nu _0| |\xi_{k+1,0}-\xi_{k,0}| \gg 1,
\end{eqnarray}
where $\nu _0 = {1  \over N }\sum_{k=1}^{N}\nu _k$ and
$ \mu _0 = {1 \over N }\sum_{k=1}^{N}\mu _k$ are, respectively, the average amplitude and
velocity of the soliton chain.
Thus, in the adiabatic approximation, one considers a chain of well-separated solitons
with amplitudes and velocities that vary slightly from their averages.
In fact, we define the following two scales:
\[ |\nu _k-\nu _0| \simeq \varepsilon_0^{1/2}, \qquad |\mu _k-\mu _0| \simeq \varepsilon_0^{1/2}, \qquad
|\xi_{k+1,0}-\xi_{k,0}| \simeq \varepsilon_0^{-1}. \]

The main idea of Karpman and Soloviev \cite{Karp} was to derive a dynamical system for the soliton
parameters which would describe the soliton interactions. In \cite{Karp} they realized the idea for
the simplest nontrivial case and described the 2-soliton interactions. The generalization of their
system to any number $N$ of solitons was proposed in \cite{PRL,PRE,PLA}. Neglecting the terms of
order higher than $\varepsilon$ allowed one to realise that in the absence of perturbations
the resulting dynamical system is  a generalization of the Toda chain to complex values of its dynamical
variables -- hence the name complex Toda chain (CTC).

Later these ideas were generalized also for the perturbed Manakov model \cite{VG1,GDM,MCS,GT,GTK1,GTK2}.

\subsection{Generalizing Karpman-Soloviev equations for $N>2$ solitons}

%{\bf check-F.mw check-Man.mw }

In this Section we will propose a simple derivation of Karpman-Soloviev equations \cite{Karp} which we have used
in our previous papers. This derivation can be easily generalized also to the Manakov model.

We start by the fact that $u_k(x,t)$ in eq.(\ref{eq:Nst}) satisfies the NLS eq.:
\begin{equation}\label{eq:nlsk}\begin{split}
i \frac{\partial u_k}{ \partial t} + \frac{1}{2} \frac{\partial u_k}{ \partial x^2 } + |u_k|^2 u_k(x,t) =0.
\end{split}\end{equation}
Next we insert the $N$-soliton train (\ref{eq:Nst}) into the perturbed NLSE (\ref{eq:nls}) and
consider it in the vicinity of $z_k$. Thus we find terms of different orders of magnitudes.
In the equation below we retain in the left hand side the leading ones depending only on $u_k$; in the
right hand side we collect the terms of next order, that take into account the nearest neighbor interactions.
The result is:
\begin{equation}\label{eq:nls2}\begin{split}
i \frac{\partial u_k}{ \partial t} + \frac{1}{2} \frac{\partial u_k}{ \partial x^2 } + |u_k|^2 u_k(x,t) =
i(R^{(0)}[u_k] +R[u_k]),
\end{split}\end{equation}
where
\begin{equation}\label{eq:nls3}\begin{split}
& R^{(0)}[u_k] e^{-i\phi_k} = i ( 2 |u_k|^2(u_{k-1} + u_{k+1}) + u_k^2(u_{k-1}^* + u_{k+1}^*)e^{-i\phi_k} \\
& \simeq \sum_{n=k\pm 1}^{}\frac{ 8\nu_0^3}{\cosh^2(z_k) \cosh(z_n)} \left( 2 e^{i(\phi_n -\phi_k)}
+e^{-i(\phi_n -\phi_k)} \right), \\
&R[u_k]e^{-i\phi_k} = (i\gamma u_k + i\beta |u_k|^2 u_k + i\eta  |u_k|^4 u_k -i V(x) u_k) e^{-i\phi_k}\\
& + i \beta \sum_{n=k\pm 1} ( 2 |u_k|^2u_{n}  +u_k^2u_{n}^*) e^{-i\phi_k}
+ i \eta \sum_{n=k\pm 1} ( 3 |u_k|^4u_{n}  +|u_k|^2 u_k^2u_{n}^*)e^{-i\phi_k}\\
&= R_0[u_k]e^{-i\phi_k}+R_1[u_k]e^{-i\phi_k},
\end{split}\end{equation}
where
\begin{equation}\label{eq:R0R1}\begin{split}
&R_0[u_k]e^{-i\phi_k}=   \frac{2i\nu_0}{\cosh(z_k)} \left( \gamma  + \frac{4  \beta \nu_0^2 }{\cosh^2 (z_k)}
+ \frac{16  \beta \nu_0^4 }{\cosh^4 (z_k)}- V(x) \right) \\
&R_1[u_k]e^{-i\phi_k} \\
&=i \beta \sum_{n=k\pm 1}^{} \frac{8\nu_0^3(2 e^{iA_{kn}} +e^{-iA_{kn}})} {\cosh^2(z_k) \cosh(z_n) }
+ i \eta \sum_{n=k\pm 1}^{} \frac{32\nu_0^5 (3 e^{iA_{kn}} +2e^{-iA_{kn}})}{\cosh^4(z_k) \cosh(z_n) }  .
\end{split}\end{equation}
where $A_{k,n}=\phi_n -\phi_k$. In the formulae above we have retained only the nearest neighbor interactions.
Thus we have neglected terms of order higher than $\varepsilon$.

The aim of the adiabatic approximation is to assume, that the soliton interactions changes
the parameters of each soliton. The idea is to derive a dynamical system of equations for these parameters.
Making this assumption we can express $\partial_t u_k $ in terms of the $t$-derivatives of the soliton
parameters as follows:
\begin{equation}\label{eq:ut-s}\begin{split}
i\frac{\partial u_k}{ \partial t }
&+ \frac{1}{2} \frac{\partial u_k}{ \partial x^2 } + |u_k|^2 u_k(x,t) =  iU(z_k) e^{i \phi_k} , \\
U(z_k)&= \frac{2\nu_k }{\cosh(z_k)}\left( (1- z_k \tanh(z_k)) \frac{1}{\nu_k} \frac{\partial \nu_k}{ \partial t} +
i \frac{ z_k}{\nu_k} \frac{\partial \mu_k}{ \partial t} \right . \\
&  \left.  - 2i (\mu_k + i \nu_k \tanh(z_k) ) \left ( \frac{\partial \xi_k}{ \partial t} - 2\mu_k \right)
+ i \left( \frac{\partial \delta_k}{ \partial t} -2(\mu_k^2 + \nu_k^2) \right) \right).
\end{split}\end{equation}

Let us introduce the functions
\begin{equation}\label{eq:Fs}\begin{aligned}
F_1(z_k) &= \frac{ 1}{2 \cosh(z_k) } , &\qquad F_2(z_k) &= \frac{ \tanh(z_k)}{2 \cosh(z_k) } , \\
F_3(z_k) &= \frac{ z}{4\nu_k^2 \cosh(z_k) }  , &\qquad F_4(z_k) &= \frac{1 -z_k \tanh(z_k)}{2\nu_k \cosh(z_k) }  ,
\end{aligned}\end{equation}
These functions are the components of the eigenfunction of $L$ and its derivative with respect to $\lambda$ of the
eigenvalue $\mu_k + i \nu_k$.

Then we conclude that the integrals:
\begin{equation}\label{eq:intNX}\begin{split}
\int_{-\infty}^{\infty} dz_k\; U(z_k) F_s (z_k) \simeq \int_{-\infty}^{\infty} dz_k\;
(R^{(0)}[u_k] + R[u_k])  F_s (z_k),
\end{split}\end{equation}
$ s= 1,\dots , 4$ must be equal up to terms of higher order of $\varepsilon$ and evaluate them explicitly.

Eq. (\ref{eq:intNX}) will be the main tool in what follows. Its
left hand sides is easily calculated in terms of the soliton parameters $t$-derivatives.
The results for $s=1,\dots , 4$ are as follows:
\begin{equation}\label{eq:NN}\begin{split}
N[u_k]  & = \frac{\partial \nu_k}{ \partial t} +
2i \nu_k \left ( \frac{\partial \delta_k}{ \partial t } -2(\mu_k^2 + \nu_k^2)  \right )
- 2i \mu_k \left ( \frac{\partial \xi_k}{ \partial t} - 2\mu_k \right ), \\
M[u_k] &= i\frac{\partial \mu_k}{ \partial t} +\frac{ 4\nu_k^2}{3}\left ( \frac{\partial \xi_k}{ \partial t}-2\mu_k \right),\\
\Xi [u_k] &=  \frac{\partial \xi_k}{ \partial t} - 2\mu_k +
i\frac{ \pi^2}{12} \frac{\partial \mu_k}{ \partial t},\\
D [u_k]  & = i \left( \frac{\partial \delta_k}{ \partial t} - 2(\mu_k^2 + \nu_k^2) \right )
+ \frac{ 1}{3} \frac{\partial \nu_k}{ \partial t} \left( 2 + \frac{\pi^2}{6} \right)
 - 2i\mu_k \left( \frac{\partial \xi_k}{ \partial t}- 2\mu_k \right).
\end{split}\end{equation}

Details about evaluating the right hand sides of the integrals (\ref{eq:intNX}) are given in the Appendix.
In what follows we will keep only terms up to the order of $\mathcal{O}(\varepsilon)$.

\subsection{Derivation of the  complex Toda chain}

In this Subsection we assume that the perturbation terms (\ref{eq:iRu}) are vanishing $R[u_k]=0$.
Then taking the real (resp. imaginary) part of eq. (\ref{eq:intNX}) with $s=1$ (resp. $s=2$) we get:
\begin{equation}\label{eq:nu-mu}\begin{split}
\frac{\partial \nu_k }{ \partial t} & = 16\nu_0^3 \re i(W_{k+1,k} + W_{k,k-1}), \\
\frac{\partial \mu_k }{ \partial t} & = \frac{ 16\nu_0^3}{3} \im  i(W_{k+1,k} - W_{k,k-1}), \\
\end{split}\end{equation}
where
\begin{equation}\label{eq:Wk}\begin{split}
W_{k,n} = e^{-|\Delta_{k,n}} e^{i(\phi_{0n} - \phi_{0k})}, \qquad \phi_{0k} = - 2\mu_0 \xi_k +\delta_k.
\end{split}\end{equation}
In addition, for definiteness we will assume that initially  $\xi_1< \xi_2 < \cdots < \xi_N$.

Next we introduce $\lambda_k = \mu_k + i \nu_k$ and after some calculations obtain:
\begin{equation}\label{eq:lamk}\begin{split}
 \frac{\partial \lambda_k}{ \partial t} = - 4\nu_0 ( e^{q_{k+1}-q_{k}} - e^{q_{k}-q_{k-1}})
\end{split}\end{equation}
where $  e^{q_{1}-q_0}=0$, $e^{q_{N+1}-q_N}=0$ and $q_k$ is given by:
\begin{equation}\label{eq:q_k}\begin{split}
 q_k &= -2\nu_0 \xi_k + k \ln (4\nu_0^2) +i (\delta_k + \delta_0 -2\mu_0 \xi_k +k\pi ), \\
\nu _0 &= {1 \over N } \sum_{s=1}^{N} \nu _s, \qquad \mu _0 = {1 \over N } \sum_{s=1}^{N} \mu _s, \qquad
\delta _0 = {1 \over N } \sum_{s=1}^{N} \delta _s.
\end{split}\end{equation}

The other two evolution equations for $d\xi_k /dt$ and  $d\delta_k /dt$ are obtained from eq. (\ref{eq:intNX})
taking the real part (resp. imaginary part) for $s=3$ (resp. $s=4$). The result is:
\begin{equation}\label{eq:xi-del-k}\begin{split}
 \frac{\partial \xi_k}{ \partial t} = 2\mu_k, \qquad \frac{\partial \delta_k}{ \partial t} = 2(\mu_k^2 +\nu_k^2),
\end{split}\end{equation}
i.e
\begin{equation}\label{eq:del-k}\begin{split}
\frac{\partial (\delta_k +\delta_0)}{ \partial t} &= 2(\mu_k^2 +\nu_k^2) + \frac{ 2}{N}\sum_{s=1}^{N}(\mu_s^2+\nu_s^2) \\
& = 4\mu_0 \mu_k + \nu_0 \nu_k + \mathcal{O}( \varepsilon).
\end{split}\end{equation}
Thus we obtain
\begin{equation}\label{eq:dqkdt}\begin{split}
\frac{\partial q_k}{ \partial t } = -4\nu_0 \lambda_k.
\end{split}\end{equation}
up to terms of the order $\mathcal{O}(\varepsilon)$;

If we now differentiate both sides of eq. (\ref{eq:dqkdt}) with respect to $t$ and make use of
eq. (\ref{eq:lamk}) we obtain:
\begin{equation}\label{eq:ctc0}\begin{split}
\frac{\partial^2 q_k}{ \partial t^2} &= 16\nu_0^2 ( e^{q_{k+1}-q_{k}} - e^{q_{k}-q_{k-1}}), \quad k=1,\dots , N-1; \\
\frac{\partial^2 q_1}{ \partial t^2} &= 16\nu_0^2  e^{q_{2}-q_{1}}, \qquad
\frac{\partial^2 q_N}{ \partial t^2} = -16\nu_0^2  e^{q_{N}-q_{N-1}},
\end{split}\end{equation}
known as the CTC because the variables $q_k$ are complex valued.

\begin{remark}\label{rem:2}
In deriving eq. (\ref{eq:xi-del-k}) we retained only the leading terms of order $\mathcal{O}(1)$ and
$\mathcal{O}(\varepsilon^{1/2})$. They contain also additional terms of the order of $\mathcal{O}(\varepsilon)$
coming from the right hand sides of eq. (\ref{eq:intNX}) which we neglect.

\end{remark}

\subsection{Derivation of the  perturbed complex Toda chain}

In order to derive the perturbed CTC (PCTC) we will take into account the
effect of perturbation terms coming from $R[u_k]$ in the right hand side of eq. (\ref{eq:intNX}).
This leads to:
\begin{equation}\label{eq:nuk0}\begin{aligned}
{d\lambda _k \over dt} &= -4\nu _0 \left(e^{q_{k+1}-q_k} - e^{q_k -q_{k-1}} \right) +
\im \mathcal{ M}_{k} +i \; \re \mathcal{ N}_{k} ,  \\
{d q_k \over dt} &=  - 4\nu_0 \lambda_k + 2i(\mu_0 +i\nu_0) \re \mathcal{X}_{k}  -i\;\im \mathcal{D}_{k}
\end{aligned}\end{equation}
where
\begin{equation}\label{eq:N-Xtil}\begin{aligned}
\mathcal{N}_k[u] &= \int_{-\infty}^{\infty} dz_k \; R[u_k] F_1(z_k), & \quad
\mathcal{M}_k[u] &= \int_{-\infty}^{\infty} dz_k \; R[u_k] F_2(z_k), \\
\mathcal{X}_k[u] &= \int_{-\infty}^{\infty} dz_k \; R[u_k] F_3(z_k), & \quad
\mathcal{D}_k[u] &= \int_{-\infty}^{\infty} dz_k \; R[u_k] F_4(z_k).
\end{aligned}\end{equation}

In fact these integrals have already been evaluated, see \cite{CGGT} and \cite{GT}.
\begin{equation}\label{eq:MD}\begin{split}
\re  \mathcal{ N}_{k} &= 2\nu_k \left( \gamma + \frac{8}{3}\beta \nu_k^2 + \frac{128}{15} \eta \nu_k^4 \right)
, \qquad \Xi_k[u]=0, \\
M_k[u] &= {\pi A\Omega^2 \over 8\nu _k  \sinh Z_k } \sin (\Omega \xi_k +\Omega _0), \\
D_k[u] &= - {\pi^2 A\Omega ^2 \over 16\nu_k^2 }{\cosh Z_k  \over \sinh ^2 Z_k } \cos (\Omega \xi_k +\Omega _0) ,
\end{split}\end{equation}
where $Z_k = \Omega \pi/(4\nu_k)$.

The final result from these calculations is the PCTC as a model
for the adiabatic $N$-soliton interactions in the presence of external potentials and
a linear and nonlinear gain/loss terms. Here for simplicity we have taken into account only the
effects of periodic potentials, for other types of potentials see \cite{124a,126a}.

\subsection{Analysis of the leading terms in the PCTC}
%%%%%%%%%%%%%%%%%%%%%%%%%%%%%%%%%%%%%%%%%%%%%%%%%%%%%%%%
In the right hand side of eq. (\ref{eq:nuk0}), even for vanishing potentials (i.e. for $A=0$) we have two types of
terms. The first type are typical for CTC, the others $\re \mathcal{N}_k$ come from nonlinear gain/loss terms, see (\ref{eq:MD}).
Let us analyze first the effects of $\re \mathcal{N}_k$ on the soliton amplitudes $\nu_k$. Note, that each $\nu_k$ is driven
separately:
\begin{equation}
\label{eq:nuk1}
\frac{d \nu_k}{d t} = 2\nu_k P(\nu_k), \qquad P_0(\nu_k) =
\gamma + \frac{8}{3}\beta \nu_k^2 + \frac{128}{15} \eta \nu_k^4.
\end{equation}
Note also, that the average amplitude $\nu_0$ will be $t$-dependent, which violates the integrability
of the PCTC.

Since typically $\nu_k \simeq 0.5$, then for generic values of $\gamma$, $\beta$ and $\eta$,  the
polynomial $\nu_k P_0(\nu_k)$
will be large (at least $\gg \mathcal{O}(\varepsilon)$) and therefore will determine the dynamics of $\nu_k$. It is easy to see,
that for generic choices of  $\gamma$, $\beta$ and $\eta$  the solutions $\nu_k(t)$ of eq. (\ref{eq:nuk1}) will either
quickly grow or quickly decay. In both cases the amplitudes will be violating the adiabaticity condition (\ref{eq:ad-ap}).

For our practical applications we will  need to restrict the parameters $\gamma$, $\beta$ and $\eta$  so that the
solutions for $\nu_k$  remain `physical' (i.e., do not blow up or vanish
quickly) for long times. In this Subsection we will list two such cases, described in our earlier paper \cite{CGGT}
for which $\eta =0$, i.e. the quintic gain/loss term is absent.

One of them, (case (A) below) corresponds to purely cubic gain/loss; the other one (case (B))
involves a combination of linear and cubic gain/loss terms. They yield, respectively:
\begin{eqnarray}
{\rm (A)} \quad \frac{d \nu_k}{d t} &=&  \frac{16\beta}{3} \nu_k^3, \label{eq:1BS1} \\
\label{eq:1BS2}
{\rm (B)} \quad \frac{d \nu_k}{d t} &=& 2\gamma \nu_k+ \frac{16\beta}{3} \nu_k^3.
\end{eqnarray}
It is evident that the gain/loss coefficients affect strongly the amplitudes of the solitons.
In fact, the above systems have the following explicit solutions:
\begin{equation}\label{eq:A+B}\begin{aligned}
&{\rm (A)} &\quad \nu_k(t) &= \frac{ \nu_k(0)}{\sqrt{1 - w_{k0}t}} , &\quad w_{k0} &= \frac{32\beta \nu_{k0}^2}{3}, \\
\\
&{\rm (B)} &\quad \nu_k(t) &= \nu_k(0) \frac{ e^{2\gamma t}}{\sqrt{1+ w_{k0}(e^{4\gamma t}-1)}},
 &\quad w_{k0} &= -\frac{8\beta \nu_{k0}^2}{\gamma^2},
%\sqrt{\frac{3\gamma }{3\gamma C_1 \exp(-4\gamma t) -8\beta }},
%
\end{aligned}\end{equation}
where the integration constants are determined by the initial soliton amplitudes.
% requiring that $P(0) = \nu_k(0)$ and $N(0) = \nu_k(0)$.
%
Other solutions with similar properties are given in Subsection 3.3.

\section{Karpman-Soloviev equations and CTC for the Manakov system}

Let us now extend the derivation of the PCTC to the Manakov system. The difference with
the scalar case is that now the we have a system of two coupled NLSEs.

\begin{equation}\label{eq:man}\begin{split}
i\vec{u}_t + \frac{1}{2} \vec{u}_{xx} +(\vec{u}^\dag \vec{u}) \vec{u}(x,t)=iR[\vec{u}].
\end{split}\end{equation}
The relevant  perturbation terms take the form:
\begin{equation}\label{eq:iRu}\begin{split}
iR[\vec{u}] &= i \left( \gamma \vec{u} + \beta (\vec{u}^\dag \vec{u}) \vec{u} + \eta
(\vec{u}^\dag \vec{u})^2 \vec{u} \right) + V(x) \vec{u}, \\
V(x) &= A \cos( \Omega x +\Omega_0).
\end{split}\end{equation}
The $N$-soliton train becomes:
\begin{equation}\label{eq:Nman}\begin{aligned}
\vec{u}(x,t=0) &= \sum_{k=1}^N \vec{u}_k(x,t=0), &\qquad \vec{u}_k(x,t) &= {2\nu_k e^{i\phi_k}\over \cosh(z_k)}\vec{n}_k  ,
\end{aligned}\end{equation}
with the same notations as in eq. (\ref{eq:Nst}. )
Thus in addition to the $4N$ standard scalar soliton parameters we have in addition $N$ 2-component
normalized polarization vectors:
\begin{equation}\label{eq:vnk}\begin{split}
\vec{n}_k =  \left(\begin{array}{c} \cos (\theta_k) e^{i\beta_k} \\ \sin (\theta_k) e^{-i\beta_k} \end{array}\right),
\qquad (\vec{n}_k^\dag \vec{n}_k) =1.
\end{split}\end{equation}
so we need to derive evolution equations also for the vectors $\vec{n}_k$.
For the unperturbed Manakov system, as well as for the perturbations with external potentials
these evolution equations had been derived in \cite{VG1,GDM,MCS,GT,GTK1,GTK2} using the variational method.
Here we will use directly the Manakov equations and
we will prove that the linear and nonlinear gain/loss do not affect the evolution of $\vec{n}_k$.

The analogs of eqs. (\ref{eq:ut-s}) and (\ref{eq:intNX}) are
\begin{equation}\label{eq:Maut-s}\begin{split}
i\frac{\partial \vec{u}_k}{ \partial t }
&+ \frac{1}{2} \frac{\partial \vec{u}_k}{ \partial x^2 } + (\vec{u}_k^\dag, \vec{u}_k) \vec{u}_k(x,t) =
i(\vec{R}^{(0)}[u_k] + \vec{R}[u_k]),
%iU(z_k) e^{i \phi_k} \vec{n}_k -\frac{2i\nu_k }{\cosh(z_k)} \frac{\partial \vec{n}_k}{ \partial t}  e^{i \phi_k} ,
\end{split}\end{equation}
and
\begin{multline}\label{eq:VintNX}
\int_{-\infty}^{\infty} dz_k\; \left( U(z_k)\vec{n}_k  -\frac{2i\nu_k }{\cosh(z_k)} \frac{\partial \vec{n}_k}{ \partial t} \right)
F_s (z_k) \\ \simeq \int_{-\infty}^{\infty} dz_k\; (\vec{R}^{(0)}[u_k] + \vec{R}[u_k])  F_s (z_k),
\end{multline}
where $U(z_k)$ is the same as in eq. (\ref{eq:ut-s}) and
\begin{equation}\label{eq:nls3M}\begin{split}
& \vec{R}^{(0)}[u_k] e^{-i\phi_k}  \simeq \sum_{n=k\pm 1}^{}\frac{ 8i\nu_0^3}{\cosh^2(z_k} \cosh(z_n) \left( (\mathcal{Y}_{kn} + \mathcal{Y}_{kn}^*)\vec{n}_k +\vec{n}_n e^{iA_{kn}} \right), \\
& \mathcal{Y}_{kn} = (\vec{n}_k^\dag, \vec{n}_n ) e^{iA_{kn}} , \qquad A_{kn} = \phi_{0,n} - \phi_{0,k}, \\
\end{split}\end{equation}
and
\begin{equation}\label{eq:vR0R1}\begin{split}
& \vec{R}[u_k]e^{-i\phi_k} = \vec{R}_0[u_k]e^{-i\phi_k} + \vec{R}_1[u_k]e^{-i\phi_k}, \\
&\vec{R}_0[u_k]e^{-i\phi_k} = \frac{ 2i\nu_0}{\cosh(z_k)} \left (\gamma + \frac{ \beta}{\cosh^2(z_k)} +
\frac{ \eta}{\cosh^4(z_k)} - V(x) \right) \vec{n}_k \\
& \vec{R}_1[u_k]e^{-i\phi_k} =i \beta \sum_{n=k\pm 1}^{} \frac{8\nu_0^3} {\cosh^2(z_k) \cosh(z_n) }
 \left( (\mathcal{Y}_{kn} + \mathcal{Y}_{kn}^*)\vec{n}_k +\vec{n}_n e^{iA_{kn}} \right) \\
&+ i \eta \sum_{n=k\pm 1}^{} \frac{32\nu_0^5 }{\cosh^4(z_k) \cosh(z_n) }
 \left( 2(\mathcal{Y}_{kn} + \mathcal{Y}_{kn}^*)\vec{n}_k +\vec{n}_n e^{iA_{kn}} \right).
\end{split}\end{equation}
where $A_{k,n}=\phi_n -\phi_k$. In the formulae above we have retained only the nearest neighbor interactions.
Thus we have neglected terms of order higher than $\varepsilon$.

Like in the previous Section  we evaluate both sides of eq. (\ref{eq:VintNX}) for $s=1, \dots , 4$.
Now the results for the left hand sides of (\ref{eq:VintNX}) are:
\begin{equation}\label{eq:ManNN}\begin{split}
&\vec{N}[u_k] =  \left( \frac{\partial \nu_k}{ \partial t} + 2i \nu_k \left ( \frac{\partial \delta_k}{ \partial t } -
2(\mu_k^2 + \nu_k^2)  \right )  -2i\mu_k \left( \frac{\partial \xi_k }{ \partial t}
- 2\mu_k \right)\right)\vec{n}_k \\
& \qquad - 2i\nu_k \frac{\partial \vec{n}_k}{ \partial t}, \\
&\vec{M}[u_k]  = i\frac{\partial \mu_k}{ \partial t}\vec{n}_k +
\frac{ 4\nu_k^2}{3} \left( \frac{\partial \xi_k}{ \partial t}- 2\mu_k \right) \vec{n}_k,\\
&\vec{\Xi} [u_k]  = \left( \frac{\partial \xi_k}{ \partial t} - 2\mu_k \right)
\vec{n}_k + i\frac{ \pi^2}{12 \nu_k^2} \frac{\partial \mu_k}{ \partial t}\vec{n}_k,
\end{split}\end{equation}
and
\begin{multline}\label{eq:Dvec}
\vec{D} [u_k]
 = i \left( \frac{\partial \delta_k}{ \partial t} - 2(\mu_k^2 + \nu_k^2 )\right)\vec{n}_k
 -2i \mu_k  \left( \frac{\partial \xi_k}{ \partial t}- 2\mu_k \right) \vec{n}_k \\
+ \frac{ 1}{3 \nu_k} \frac{\partial \nu_k}{ \partial t} \left( 2 + \frac{\pi^2}{6} \right)\vec{n}_k
-i \frac{\partial \vec{n}_k}{ \partial t}.
\end{multline}
Note that the second and the third of the equations in (\ref{eq:ManNN}) do not contain $\partial_t \vec{n}_k $.

\subsection{Generalizing the CTC to the Manakov model }

Here we will generalize the CTC to the Manakov model. To this end in this
Subsection we assume that $\vec{R}[u_k] =0$. Like in the scalar case, we multiply the right hand side of
eq. (\ref{eq:Maut-s}) by $F_s(z_k)$, integrate over $z_k$ neglecting terms of order higher than $\mathcal{O}(\varepsilon)$.
The integrals that we need to calculate are the same as for the scalar case, see the Appendix.
Thus for the right hand sides of the integrals (\ref{eq:VintNX}) we find:
\begin{equation}\label{eq:N-X-vec}\begin{aligned}
\vec{\mathcal{N}}_k[u] & = 16i \nu_0^3 \sum_{n=k\pm 1}^{} e^{-|\Delta_{k,n}|}\left( (\mathcal{Y}_{kn}
+ \mathcal{Y}_{kn}^*) \vec{n}_k + \vec{n}_n e^{iA_{kn}} \right ), \\
\vec{\mathcal{M}}_k[u] & = \frac{ 16i \nu_0^3}{3} \sum_{n=k\pm 1}^{} \epsilon_n e^{-|\Delta_{k,n}|}\left( (\mathcal{Y}_{kn}
+ \mathcal{Y}_{kn}^*) \vec{n}_k + \vec{n}_n e^{iA_{kn}} \right ), \\
\vec{\mathcal{X}}_k[u] & =  4i \nu_0 \sum_{n=k\pm 1}^{} \epsilon_n e^{-|\Delta_{k,n}|}\left( (\mathcal{Y}_{kn}
+ \mathcal{Y}_{kn}^*) \vec{n}_k + \vec{n}_n e^{iA_{kn}} \right ), \\
\vec{\mathcal{D}}_k[u] & =  8i \nu_0^2 \sum_{n=k\pm 1}^{}  e^{-|\Delta_{k,n}|}\left( (\mathcal{Y}_{kn}
+ \mathcal{Y}_{kn}^*) \vec{n}_k + \vec{n}_n e^{iA_{kn}} \right ),
\end{aligned}\end{equation}
where $\epsilon_{k+1}=1$, $\epsilon_{k+1}=-1$.

First we take the scalar products of both sides of eqs. (\ref{eq:N-X-vec}) with $ \langle \vec{n}_k^\dag|$. this
gives us the chance to obtain the evolution of the $\nu_k$, $\mu_k$, $\xi_k$ and $\delta_k$. Comparing
eqs. (\ref{eq:ManNN}), (\ref{eq:Dvec}) and (\ref{eq:VintNX}) we obtain:
\begin{equation}\label{eq:dnu-xit}\begin{aligned}
\frac{\partial \nu_k }{ \partial t} &= \re \langle \vec{n}_k^\dag , \vec{\mathcal{N}}_k[u] \rangle, & \quad
\frac{\partial \xi_k }{ \partial t} -2\mu_k &= \re \langle \vec{n}_k^\dag , \vec{\mathcal{X}}_k[u] \rangle, \\
\frac{\partial \mu_k }{ \partial t} &= \im \langle \vec{n}_k^\dag , \vec{\mathcal{M}}_k[u] \rangle, & \quad
\frac{\partial \delta_k }{ \partial t} -2(\mu_k^2+\nu_k^2) &= \im  \langle \vec{n}_k^\dag , \vec{\mathcal{D}}_k[u] \rangle .
\end{aligned}\end{equation}

In addition from eq. (\ref{eq:Dvec}) and the last of the eqs. (\ref{eq:N-X-vec}) we find the evolution of the
polarization vectors:
\begin{equation}\label{eq:vecNk}\begin{split}
-i \frac{\partial \vec{n}_k}{ \partial t} &= \vec{\mathcal{D}}_k[u] + i \left( \frac{\partial \delta_k}{ \partial t}
- 2(\mu_k^2 + \nu_k^2) +2\mu_k \left( \frac{\partial \xi_k}{ \partial t} - 2\mu_k \right) \right ) \vec{n}_k \\
& \qquad -\frac{1}{3\nu_k} \left( 2 + \frac{\pi^2}{6} \right) \frac{\partial \nu_k}{ \partial t} \vec{n}_k .
\end{split}\end{equation}

\begin{remark}\label{rem:1}
The right hand side of eq. (\ref{eq:vecNk}) is of the order of $\mathcal{O}(\varepsilon)$, which will be enough for
our future analysis.
\end{remark}

Next we evaluate $d\mu_k /dt$ and  $d\nu_k /dt$ obtaining:
\begin{equation}\label{eq:dnuk}\begin{split}
\frac{\partial \mu_k}{ \partial t} & = -16\nu_0^3 \sum_{n=k\pm 1}^{} \epsilon_n e^{-|\Delta_{kn}|} \re \mathcal{Y}_{k,n}, \\
\frac{\partial \nu_k}{ \partial t} &= 16\nu_0^3 \sum_{n=k\pm 1}^{}  e^{-|\Delta_{kn}|} \im  \mathcal{Y}_{k,n},
\end{split}\end{equation}
Thus after some calculations we find \cite{PRE}:
\begin{equation}\label{eq:dlat}\begin{split}
\frac{\partial (\mu_k + i \nu_k)}{ \partial t} = -4\nu_0 (e^{q_{k+1} - q_{k}} \langle \vec{n}_{k+1}^\dag , \vec{n}_{k} \rangle
- e^{q_{k} - q_{k-1}} \langle \vec{n}_{k}^\dag , \vec{n}_{k-1} \rangle )
\end{split}\end{equation}
with the same choice for $q_k$ as in (\ref{eq:q_k}).

It is easy to check that the evolution equations for $\xi_k$ and $\delta_k$ remain the same as for
the scalar case, see eq. (\ref{eq:xi-del-k}). Thus we again get:
\begin{equation}\label{eq:dq-k}\begin{split}
\frac{\partial q_k}{ \partial t} = - 4\nu_0 \lambda_k,
\end{split}\end{equation}
and the generalized CTC (GCTC) for the Manakov model becomes:
\begin{equation}\label{eq:ctcMa}\begin{split}
\frac{\partial^2 q_1}{ \partial t^2} &= 16\nu_0^2 e^{q_{2} - q_{1}} \langle \vec{n}_{1}^\dag ,\vec{n}_{1} \rangle \\
\frac{\partial^2 q_k}{ \partial t^2} &= 16\nu_0^2 (e^{q_{k+1} - q_{k}} \langle \vec{n}_{k+1}^\dag , \vec{n}_{k} \rangle
- e^{q_{k} - q_{k-1}} \langle \vec{n}_{k}^\dag , \vec{n}_{k-1} \rangle ) , \\
\frac{\partial^2 q_k}{ \partial t^2} &= -16\nu_0^2 e^{q_{N} - q_{N-1}} \langle \vec{n}_{N}^\dag , \vec{n}_{N-1} \rangle .
\end{split}\end{equation}

\begin{remark}\label{rem:3}
Generically speaking the system (\ref{eq:ctcMa}) must be extended with the equations (\ref{eq:vecNk})
describing the evolution of the polarization vectors. Note however, that eq. (\ref{eq:ctcMa}) depends on the
polarization vectors only through the scalar products $\langle \vec{n}_{k}^\dag , \vec{n}_{k-1} \rangle$
which in view of eq. (\ref{eq:vecNk}) evolve very slowly and differ from their initial values by terms of the
order $\mathcal{O}(\varepsilon)$. Since the factors $e^{q_{k+1} - q_{k}}$ are also of the order of
$\mathcal{O}(\varepsilon)$ then we can replace the scalar products $\langle \vec{n}_{k}^\dag , \vec{n}_{k-1} \rangle$
by their initial values at $t=0$.

\end{remark}

%%%%%%%%%%%%%%%%%%%%%%%%%%%%%%%%%%%%%%%%%%%%%%%%%%%%%%%%

\subsection{Derivation of the  PCTC for the Manakov model}

The derivation of PCTC for the Manakov model is rather similar to the one for the scalar case.
Indeed, taking into account the effect of perturbation terms $\vec{R}[u_k]$ to the right hand side of eq. (\ref{eq:VintNX}).
leads to:
\begin{equation}\label{eq:nuk0M}\begin{aligned}
{d\lambda _k \over dt} &= -4\nu _0 \left(e^{q_{k+1}-q_k}\langle \vec{n}_{k+1}^\dag , \vec{n}_{k} \rangle
 - e^{q_k -q_{k-1}}\langle \vec{n}_{k}^\dag , \vec{n}_{k-1} \rangle \right) +
\im \tilde{\mathcal{ M}}_{k} +i \; \re \tilde{\mathcal{ N}}_{k} ,  \\
{d q_k \over dt} &=  - 4\nu_0 \lambda_k + 2i(\mu_0 +i\nu_0) \re \tilde{\mathcal{X}}_{k}  -i\;\im
\tilde{\mathcal{D}}_{k}
\end{aligned}\end{equation}
where
\begin{equation}\label{eq:N-XtilM}\begin{aligned}
\tilde{\mathcal{N}}_k[u] &= \int_{-\infty}^{\infty} dz_k \; \langle \vec{n}_k^\dag, \vec{R}[u_k]\rangle F_1(z_k), & \;
\tilde{\mathcal{M}}_k[u] &= \int_{-\infty}^{\infty} dz_k \;  \langle \vec{n}_k^\dag, \vec{R}[u_k] F_2(z_k), \\
\tilde{\mathcal{X}}_k[u] &= \int_{-\infty}^{\infty} dz_k \;  \langle \vec{n}_k^\dag, \vec{R}[u_k] F_3(z_k), & \;
\tilde{\mathcal{D}}_k[u] &= \int_{-\infty}^{\infty} dz_k \;  \langle \vec{n}_k^\dag, \vec{R}[u_k] F_4(z_k).
\end{aligned}\end{equation}

In fact these integrals have already been evaluated, see \cite{CGGT} and \cite{GT} and Section 2 above.
\begin{equation}\label{eq:MDM}\begin{split}
\re  \tilde{\mathcal{ N}}_{k} &= 2\nu_k \left( \gamma + \frac{8}{3}\beta \nu_k^2 + \frac{128}{15} \eta \nu_k^4 \right)
, \qquad \tilde{\mathcal{X}}_k[u]=0, \\
\tilde{\mathcal{M}}_k[u] &= {\pi A\Omega^2 \over 8\nu _k  \sinh Z_k } \sin (\Omega \xi_k +\Omega _0), \\
\tilde{\mathcal{D}}_k[u] &= - {\pi^2 A\Omega ^2 \over 16\nu_k^2 }{\cosh Z_k  \over \sinh ^2 Z_k } \cos (\Omega \xi_k +\Omega _0) ,
\end{split}\end{equation}
where $Z_k = \Omega \pi/(4\nu_k)$.
Again we have taken into account only the
effects of periodic potentials, for other types of potentials see \cite{GT,124a,126a}.

%%%%%%%%%%%%%%%%%%%%%%%%%%%%%
\subsection{Balancing the linear and nonlinear gain/loss terms}

The arguments given in Subsection 2.5 for the scalar NLSE case are quite valid also for the Manakov case.
Again the dynamics of the soliton amplitudes $\nu_k$ is determined by two types of terms. The first one
$\re \tilde{ \mathcal{N}}_k$ is typical for the CTC of Manakov type. The second type of terms for vanishing potentials
(i.e. for $A=0$) is due to the gain/loss terms:
\begin{equation}\label{eq:dnukM}\begin{split}
\frac{\partial \nu_k}{ \partial t} = 2\nu_k \left ( \gamma + \frac{8\beta}{3} \nu_k^2 + \frac{128 \eta}{15} \nu_k^4 \right ).
\end{split}\end{equation}
Let us multiply eq. (\ref{eq:dnukM}) by $\nu_k$ and introduce $z(t) = \nu_k^2(t)$, dropping for brevity the index $k$:
\begin{equation}\label{eq:dz1}\begin{split}
\frac{\partial z}{ \partial t} = 4z P(z), \qquad P(z)=\left ( \gamma + \frac{8\beta}{3} z + \frac{128\eta }{15} z^2 \right ).
\end{split}\end{equation}
Since the initial value of $\nu_k \simeq 0.5$, then as a typical initial value for $z(t)$  we will take $z(t) \simeq 0.25$.

Following the discussion in Subsection 2.5, compatibility with the adiabatic approximation requires that
the solution of (\ref{eq:dz1}) $z(t)$ must be taking values in a small interval around $z_0$. Below we obtain such solution assuming
that $P(z)$ has two positive roots $z_{1,2} = z_0 \pm a$, $a\ll 1$.

Our hypothesis is that the evolution of $z(t)$ will be compatible with the adiabatic approximation if
the right hand side of eq. (\ref{eq:dnukM}), or equivalently, of  (\ref{eq:dz1}) is of the same order of magnitude
as $\mathcal{O}(\varepsilon)$.  It is not difficult to check that if we choose
\begin{equation}\label{eq:ebgam}\begin{split}
 \eta = \frac{15}{128}, \qquad \beta = -0.1875, \qquad \gamma =0.062499,
\end{split}\end{equation}
$P(z)$ will have as roots $z_1 =0.249$ and $z_1 =0.251$.

Let us now try to solve the ODE (\ref{eq:dz1}) with the initial condition $z(0)=0.25$. We have not been able to get
explicit solution for (\ref{eq:dz1}). However we slightly simplified the equation solving
\begin{equation}\label{eq:dz2}\begin{split}
 \frac{\partial z}{ \partial t} = 4z_0 (z - z_0-a) (z-z_0+a), \qquad z_0 = 0.25,
\end{split}\end{equation}
where $a$ is a small parameter of the order of $\mathcal{O}(\varepsilon)$. Obviously the solution of (\ref{eq:dz2})
\begin{equation}\label{eq:zt}\begin{split}
 z(t)= 0.25 - a \tanh (at) , \qquad \mbox{or} \qquad \nu_k(t) = \sqrt{0.25 - a \tanh (at)}
\end{split}\end{equation}
can be used as a good approximation to exact solution of (\ref{eq:dz1}).
\begin{figure}[h!]
\centerline{\includegraphics[width=0.85\textwidth]{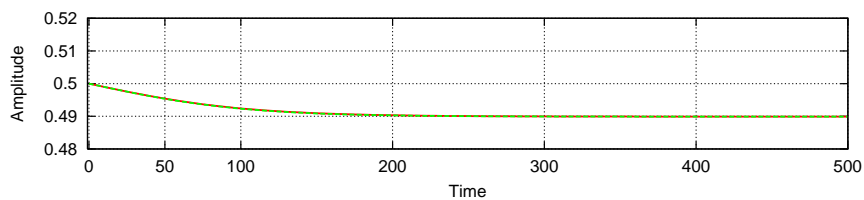}} %{dnuk-01.eps}}
%\vspace{-0.1in}
\caption{The shape of $\nu_k(t)$, $\nu_k(0)=0.50$ obtained from the solution of eq. (\ref{eq:zt}) with $a=0.01$: analytic approximation (red); numerical solution (green)}
\label{fig1}
\end{figure}
It is easy to see that this solution for all times is very close to the initial value of $\nu_k=0.5$.
The latter is confirmed by the Runge-Kutta of fourth order numerical solution of the above
initial value problem. The relative difference between both predictions does
not exceed $10^{-4}\%$ (Fig. 1).

It is not difficult to get other approximate solutions to eq. (\ref{eq:dz2}) compatible
with the adiabatic approximations. Indeed, one can slightly modify the initial condition $z_0$, by, e.g.
$z_0 =\nu_k^2(0)$, or slightly modify the positions of the zeroes of $P(z)$.

%%%%%%%%%%%%%%%%%%%%%%%%%%%%%%%%%%%%%%%%%%%%%%%%%%%%%%%%

\section{Numerical results}
\begin{figure}[h!]
\centerline{\includegraphics[width=0.9\textwidth]{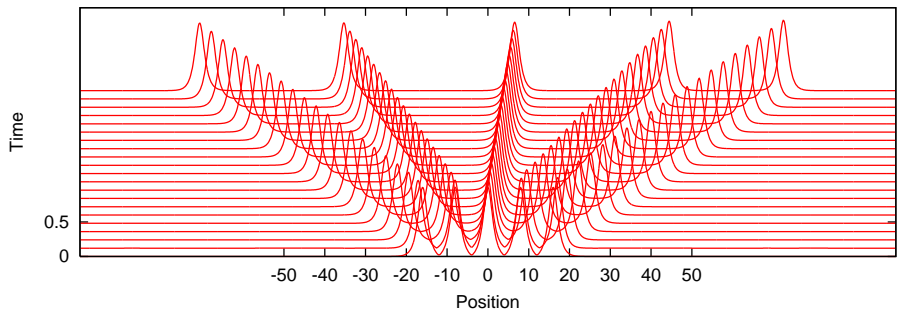}}
%\vspace{-0.1in}
\caption{5-soliton train -- Homogeneous scalar NLSE: constant $\Delta\delta=0.8\pi$ with $\delta_1=0.8\pi$; $\nu_{1,2,3,4,5}=0.5$}
\label{fig_r1a}
\end{figure}

\begin{figure}[h!]
\centerline{\includegraphics[width=0.45\textwidth]{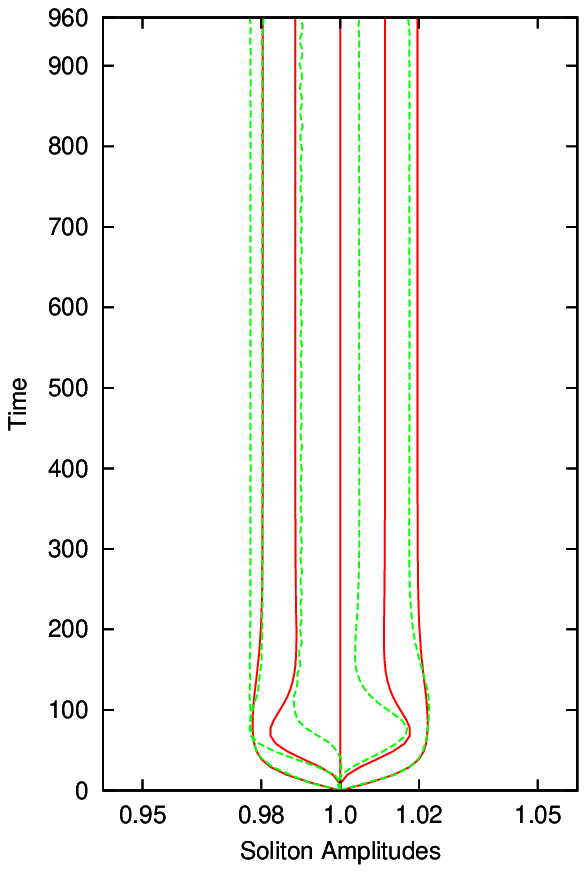}\includegraphics[width=0.45\textwidth]{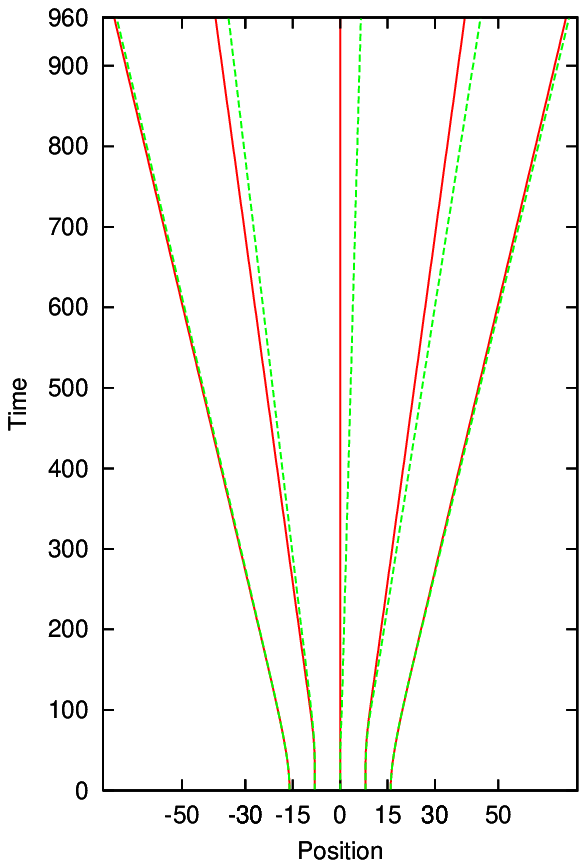}}
%\vspace{-0.1in}
\caption{Homogeneous NLSE (green) against CTC (red): constant $\Delta\delta=0.8\pi$ with $\delta_1=0.8\pi$; $\nu_{1,2,3,4,5}=0.5$}
\label{fig_r1b}
\end{figure}

To test the reliability of the derived PCTC we solve numerically CTC by Runge-Kutta method of fourth order and the nonlinear scalar NLSE by conservative fully implicit finite-difference method \cite{TodChri07} and compare the results. An object of investigation are 5-soliton trains in adiabatic approximation. The complete investigation is a hard task besides PCTC is $4N$ ($N=5$) while the GCTC -- $6N$ manifold. In all our computations we consider uniformly placed at distance $r_0=8$ solitons or soliton envelopes and vary the initial phase differences $\Delta\delta_k=\delta_{k+1}-\delta_k$, initial velocities $\nu_k$, polarization vectors $\vec n_k$, $k=1,...,5$. So, the initial positions of the soliton centers are chosen to be $\xi_{1,5}=\pm16$, $\xi_{2,4}=\pm8$, $\xi_3=0$. In Figs.~\ref{fig_r1a} and \ref{fig_r1b} the dynamics of the chosen soliton configuration as well as the trajectories of the soliton centers and the magnitudes of the soliton amplitudes are plotted. Though for the concrete set of parameters the scale $\varepsilon\simeq 10^{-2}$ and the respective ``adiabatic'' time is $1/\varepsilon$ our observations show that the comparison between CTC and NLSE is very good for considerably large time.  Similar tests to compare GCTC and Manakov system (MS) are conducted. We aim to check and specify the parameter ranges where the adiabaticity of the solutions holds.

\begin{figure}[h!]
\centerline{\includegraphics[width=0.9\textwidth]{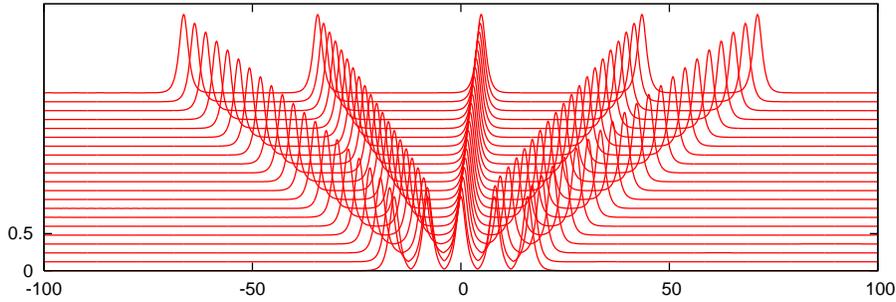}}
%\vspace{-0.1in}
\caption{5-soliton train -- Perturbed scalar NLSE: $\gamma=10^{-3}$, $\beta=-1.49\gamma$, constant $\Delta\delta=0.8\pi$ with $\delta_1=0.8\pi$; $\nu_{1,2,3,4,5}=0.5$}
\label{fig_r2a}
\end{figure}

\begin{figure}[h!]
\centerline{\includegraphics[width=0.45\textwidth]{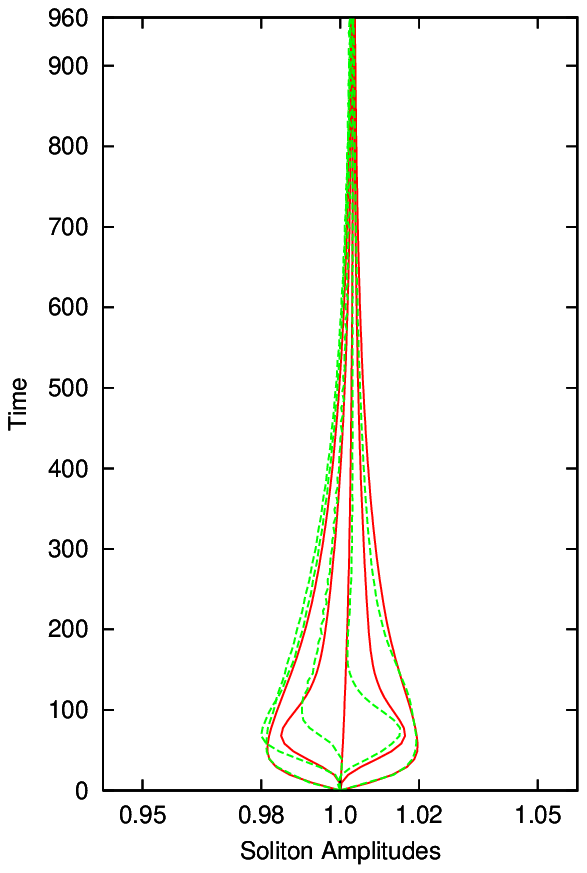}\includegraphics[width=0.45\textwidth]{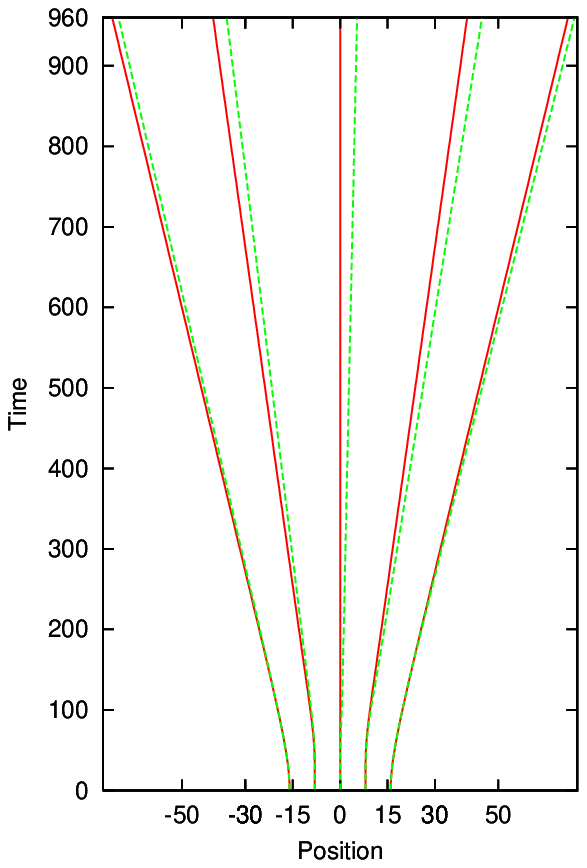}}
%\vspace{-0.1in}
\caption{Perturbed scalar NLSE (green) against PCTC (red): $\gamma=10^{-3}$, $\beta=-1.49\gamma$, constant $\Delta\delta=0.8\pi$ with $\delta_1=0.8\pi$; $\nu_{1,2,3,4,5}=0.5$}
\label{fig_r2b}
\end{figure}

\begin{figure}[h!]
\centerline{\includegraphics[width=0.45\textwidth]{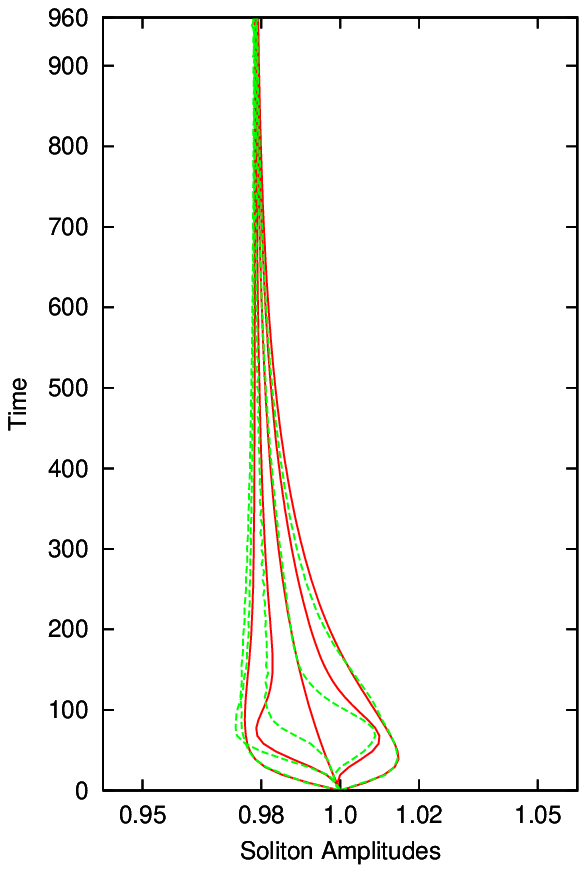}\includegraphics[width=0.45\textwidth]{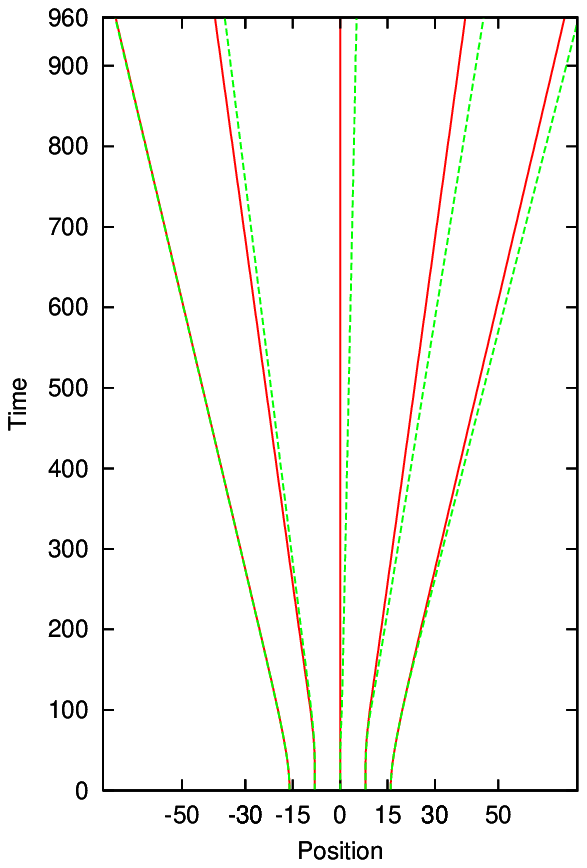}}
%\vspace{-0.1in}
\caption{Perturbed scalar NLSE (green) against PCTC (red): $\gamma=10^{-3}$, $\beta=-1.5\gamma$, $\eta=-10^{-4}$, constant $\Delta\delta=0.8\pi$ with $\delta_1=0.8\pi$; $\nu_{1,2,3,4,5}=0.5$}
\label{fig_r2c}
\end{figure}

The PCTC as well as the perturbed NLSE and perturbed MS are not integrable. After the obtained results for 1- and 2-soliton solution \cite{CGGT} the stability for NLSE with gain/loss is possible for small linear and cubic terms with opposite signs. An amplitude balance is attained when  $\beta \simeq -1.5\gamma$. Larger values of the relation lead to an infinite energy (blow-up) while the smaller ones -- to selfdispersion.  If the gain/loss (the pair ($\gamma, \beta)\neq (0,0)$) and $\eta=0$, then for $\nu_k=\sqrt{-3\gamma/8\beta}$, $k=1,2,3,4,5$ the reduced polynomial of second degree  $P(\nu_k)\equiv 0$. We established that when the coefficient $\eta\neq0$ also and the amplitudes $\nu_k$ are close to the real zeros of the bisquare polynomial $P(\nu_k)$ and an appropriate set  of the triple ($\gamma$, $\beta$, $\eta$) the perturbation term in PCTC vanishes, i.e. obtain again the original CTC with a Lax presentation and an analytical estimate of the asymptotics. Otherwise it turned out that the adiabaticity holds even a 5-degree nonlinear is present provided $\eta$, $\gamma$, $\beta$ to be small enough and latter two with opposite signs. These properties are illustrated in Figs.~\ref{fig_r2a}, \ref{fig_r2b}, \ref{fig_r2c} where the initial amplitudes are equal,
%\begin{figure}[h!]
%\centerline{\includegraphics[width=0.9\textwidth]{Scalarclasheps_gbe.eps}}
%\vspace{-0.1in}
%\caption{5-soliton train -- Perturbed scalar Schr\"odinger equation:  $\gamma=10^{-3}$, $\beta=-1.5\gamma$, $\eta=-10^{-4}$, $\Delta\delta=0.8\pi$; $\nu_{1,2,3,4,5}=0.5$; $\xi_{5,1}=\pm16$; $\xi_{3,2}=\pm8$; $\xi_3=0$.}
%\label{fig1}
%\end{figure}
$\nu_k=0.5$, and $\gamma=10^{-3}$, $\beta=-1.49\gamma$, $\eta=10^{-3}$. Comparing Figs.~\ref{fig_r2b} and \ref{fig_r2c}
one can clearly notices that the plugging on a perturbing nonlinearity of fifth degree strives to deform the trajectories
to the left or right depending on the $\eta$ sign. At that time the neither adiabiticity nor the asymptotical behavior of
the trajectories are changed. It is seen that the difference between the curves in Fig.~ \ref{fig_r1a} and \ref{fig_r2a} is
 slightly small. We conclude that the choice of an appropriate set of the coefficients in polynomial $P(\nu_k)$ including
 their signs violate the CTC integrability but not the adiabaticity including the asymptotic regime of the solitons.

\begin{figure}[h!]
\centerline{\includegraphics[width=0.9\textwidth]{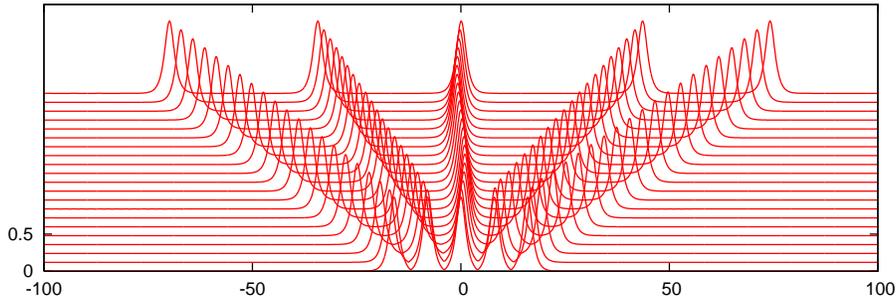}}
%\vspace{-0.1in}
\caption{5-soliton train -- Perturbed scalar NLSE:  $\gamma=10^{-3}$, $\beta=-1.49\gamma$, $\eta=-10^{-4}$, $A=-10^{-4}$, $\Omega=\pi/4$, constant $\Delta\delta=0.8\pi$ with $\delta_1=0.8\pi$; $\nu_{1,2,3,4,5}=0.5$}
\label{fig_r3a}
\end{figure}

\begin{figure}[h!]
\centerline{\includegraphics[width=0.9\textwidth]{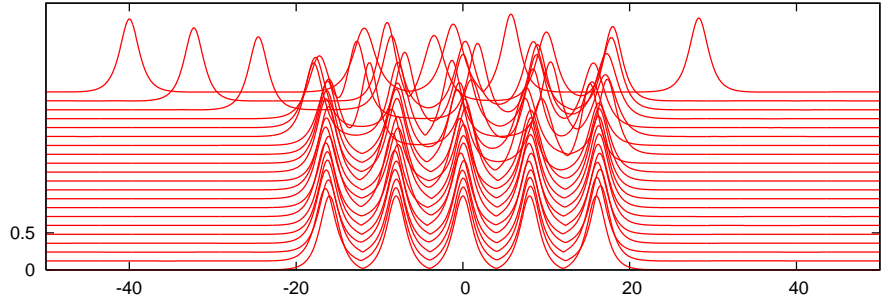}}
%\vspace{-0.1in}
\caption{5-soliton train -- Perturbed scalar NLSE: $\gamma=10^{-3}$, $\beta=-1.5\gamma$, $\eta=-10^{-4}$, $A=-0.0075$, $\Omega=\pi/4$, constant $\Delta\delta=0.8\pi$ with $\delta_1=0.8\pi$; $\nu_{1,2,3,4,5}=0.5$}
\label{fig_r3b}
\end{figure}

\begin{figure}[h!]
\centerline{\includegraphics[width=0.45\textwidth]{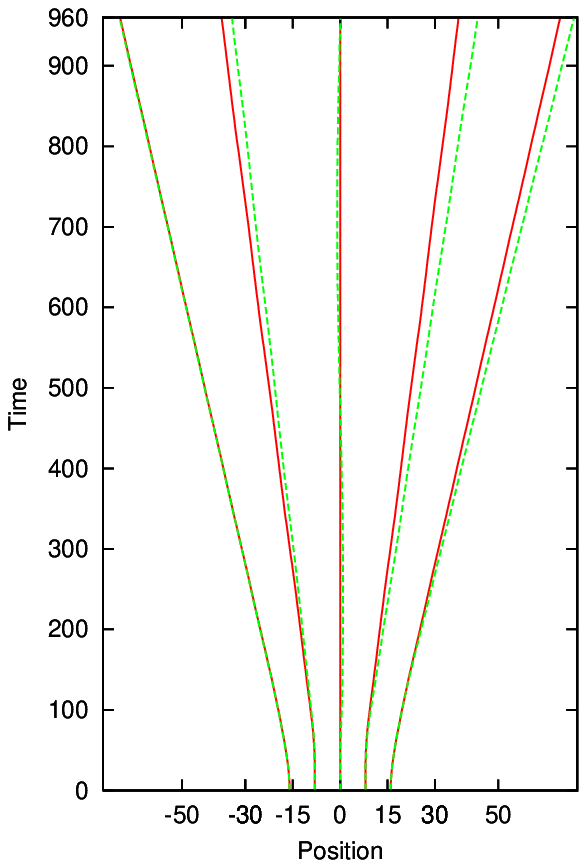}\includegraphics[width=0.45\textwidth]{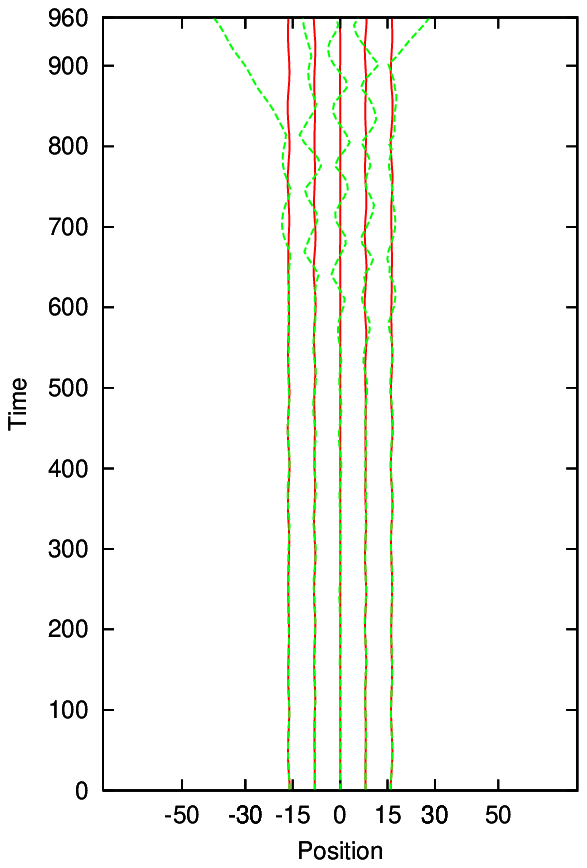}}
%\vspace{-0.1in}
\caption{Perturbed scalar NLSE (green) against PCTC (red): $\gamma=10^{-3}$, $\beta=-1.49\gamma$, $\eta=-10^{-4}$, $A=-10^{-4}$ (left), $A=0.0075$ (right), $\Omega = \pi/4$, constant $\Delta\delta=0.8\pi$ with $\delta_1=0.8\pi$; $\nu_{1,2,3,4,5}=0.5$}
\label{fig_r3c}
\end{figure}

In contrast to the linear gain/loss and the nonlinear perturbations of higher degree the periodic potential is able to change
 the asymptotic behavior \cite{GT}. Depending on the sign of amplitude $A$ it is possible a transition from a free
 asymptotic regime (FAR) to a bound state regime (BSR) or a mixed state regime (MSR) and vice versa. In the next 3 graphs
 are given results for small magnitude of $A$, when a periodic oscillation of the soliton centers appears accompanied by a
 slight decrease of the phase velocities (Fig.~\ref{fig_r3a}), and a change of the asymptotic behavior, when amplitude
 $A$ 
%\begin{figure}[h!]
%\centerline{\includegraphics[width=0.45\textwidth]{amplitudes5s_gbea.eps}\includegraphics[width=0.45\textwidth]{amplitudes5s_gbeacr.eps}}
%\vspace{-0.1in}
%\caption{Perturbed scalar Schr\"odinger equation (green) against PCTC (red): $\gamma=10^{-3}$, $\beta=-1.5\gamma$, $\eta=-10^{-4}$, $A=-10^{-4}$, $\Omega=\pi/4$, $\Delta\delta=0.8\pi$; $\nu_{1,2,3,4,5}=0.5$; $\xi_{5,1}=\pm16$; $\xi_{3,2}=\pm8$; $\xi_3=0$.}
%\label{fig1}
%\end{figure}
reaches a critical value (in the concrete case $A_{\rm crit}=-0.0075$). At that value the comparison with scalar NLSE is
excellent up to time $t \simeq 600$ (see Figs.~\ref{fig_r3b}, \ref{fig_r3c}). Yet, the further growth of $A$ still keeps the
adiabaticity at least to time $t\sim 1000$. A positive amplitude $A$ and initial soliton amplitudes $\nu_k\simeq 0.5$,
$k=1,2,3,4,5$ result an opposite effect, i.e. from BSR or MAR to FAR.
%\begin{figure}[h!]
%\centerline{\includegraphics[width=0.9\textwidth]{Scalarclasheps_hacr.eps}}
%\vspace{-0.1in}
%\caption{5-soliton train -- Perturbed scalar Schr\"odinger equation: $A=-0.075$, $\Omega=-10^{-4}$, $\Delta\delta=0.8\pi$; $\nu_{1,2,3,4,5}=0.5$; $\xi_{5,1}=\pm16$; $\xi_{3,2}=\pm8$; $\xi_3=0$.}
%\label{fig1}
%\end{figure}
%\begin{figure}[h!]
%\centerline{\includegraphics[width=0.45\textwidth]{amplitudes5s_hacr.eps}\includegraphics[width=0.45\textwidth]{trajects5s_hacr.eps}}
%\vspace{-0.1in}
%\caption{Perturbed scalar Schr\"odinger equation (right) against PCTC (left): $\gamma=\beta=\eta=0$, $\Omega=-10^{-4}$, $\Delta\delta=0.8\pi$; $\nu_{1,2,3,4,5}=0.5$; $\xi_{5,1}=\pm16$; $\xi_{3,2}=\pm8$; $\xi_3=0$.}
%\label{fig1}
%\end{figure}

\begin{figure}[h!]
\centerline{\includegraphics[width=0.9\textwidth]{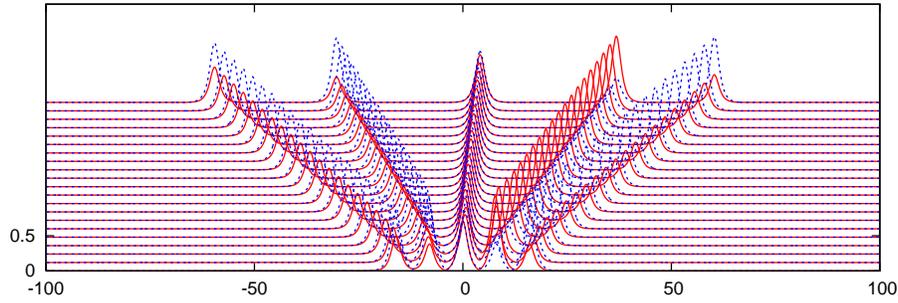}}
%\vspace{-0.1in}
\caption{Five freely propagating solitons -- Perturbed two-component MS with linear gain $\gamma=10^{-4}$ and periodic external potential: $A=-10^{-4}$, $\Omega=\pi/4$, constant $\Delta\delta=0.8\pi$ with $\delta_1=0.8\pi$; $\nu_{1,2,3,4,5}=0.5$}
\label{fig_r4a}
\end{figure}

\begin{figure}[h!]
\centerline{\includegraphics[width=0.9\textwidth]{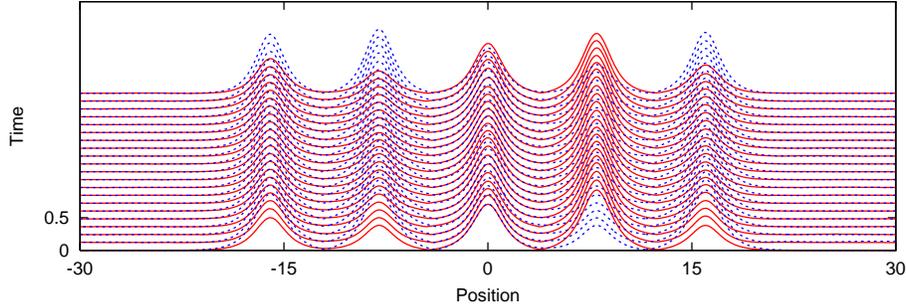}}
%\vspace{-0.1in}
\caption{Five solitons -- bound state regime due to the periodic potential. Perturbed two-component MS with linear gain $\gamma=10^{-4}$ and periodic external potential: $A=-0.075$, $\Omega=\pi/4$, constant $\Delta\delta=0.8\pi$ with $\delta_1=0.8\pi$; $\nu_{1,2,3,4,5}=0.5$}
\label{fig_r4b}
\end{figure}

\begin{figure}[h!]
\centerline{\includegraphics[width=0.45\textwidth]{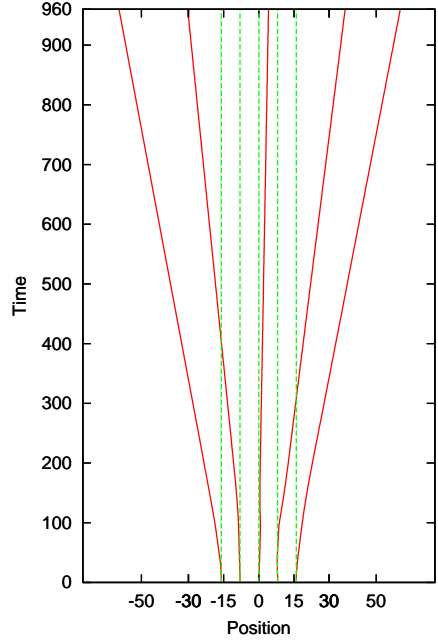}}
%\vspace{-0.1in}
\caption{5-soliton train -- Perturbed by a periodic potential two-component MS with linear gain $\gamma=10^{-4}$, $A=-10^{-4}$ (red) and $A=-0.075$ (green), $\Omega=\pi/4$, constant $\Delta\delta=0.8\pi$ with $\delta_1=0.8\pi$; $\nu_{1,2,3,4,5}=0.5$; polarization angles $\theta_1=\frac{8\pi}{24}$, $\theta_2=\frac{15\pi}{24}$, $\theta_3=\frac{18\pi}{24}$, $\theta_4=\frac{27\pi}{24}$, $\theta_5=\frac{8\pi}{24}$}
\label{fig_r4c}
\end{figure}

Further we consider two examples of PCTC for two-component MS with gain/loss and external periodic potential. One more equation is needed in this case -- for the polarization vectors. As we commented above it can be neglected in adiabatic approximation and the polarization vectors (determined by angles $\theta_k$ and $\beta_k=0$, $k=1,2,3,4,5$) to be considered as constant in the time. The latter means that PCTC of the perturbed MS is the same like those in the scalar case. In Figs.~\ref{fig_r4a}, \ref{fig_r4b}, and \ref{fig_r4c} is present the effect of the periodic potential for small linear gain ($\gamma=10^{-4}$) and two magnitudes of the periodic amplitude $A$. Similar to the scalar case  for large enough values of $A$ occurs a change of the asymptotics of soliton envelopes  from FAR to BSR. The adiabatics keeps, the soliton envelopes, too. There is no qualitative difference in the individual soliton amplitudes and their dynamic behavior is similar to the scalar case.

In the next graph (Fig.~\ref{fig_r5a}) is considered  a 5-soliton configuration again with gain/loss perturbation ($\beta=-1.49\gamma$) but with different initial soliton amplitudes and phase shifts. The corresponding asymptotic regime is MAR. Superposing a periodic potential one observes a change of the asymptotic behavior keeping the adiabaticity. The transition depends on the magnitude of periodic amplitude $A$ (Figs.~\ref{fig_r5b} and \ref{fig_r5c}). The corresponding influence over the trajectories of the soliton envelopes is illustrated in Fig.~{\ref{fig_r5d}. All the results are verified and compared with the finite-difference implementation of the perturbed MS.
\begin{figure}[h!]
\centerline{\includegraphics[width=0.9\textwidth]{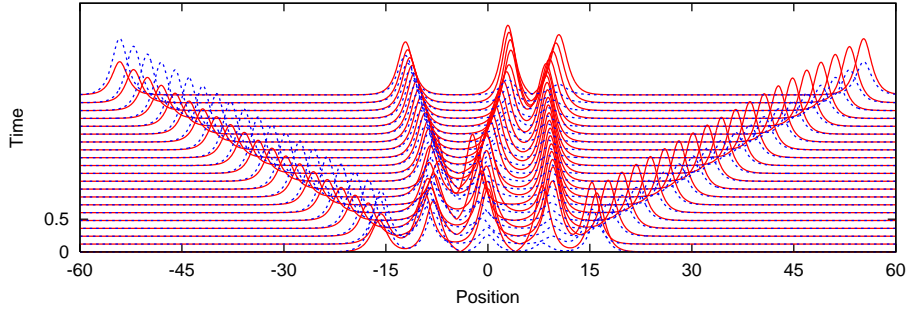}}
%\vspace{-0.1in}
\caption{Mixed asymptotic regime of 5-soliton envelopes -- Perturbed two-component MS with linear gain $\gamma=10^{-3}$ and cubic loss $\beta=-1.49\gamma$, $\delta_{1,2,4,5}=0$, $\delta_3=\pi$; $\nu_3=0.5$, $\nu_{4,2}=\nu_3\pm0.01$, $\nu_{5,1}=\nu_3\pm0.02$}
\label{fig_r5a}
\end{figure}
\begin{figure}[h!]
\centerline{\includegraphics[width=0.9\textwidth]{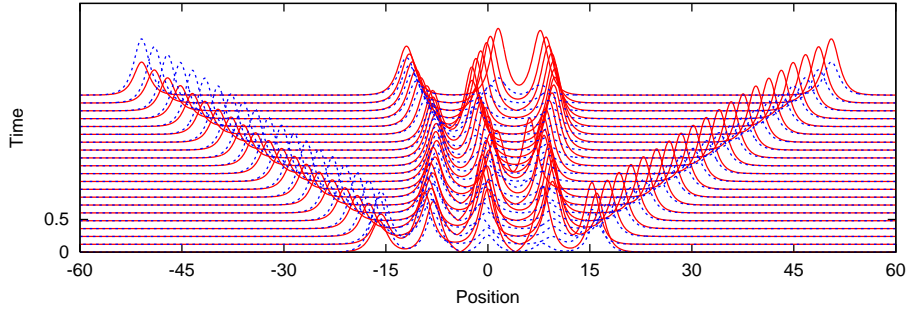}}
%\vspace{-0.1in}
\caption{Mixed asymptotic regime of 5-soliton envelopes. Perturbed two-component MS with linear gain $\gamma=10^{-4}$, cubic loss $\beta=-1.49\gamma$  and subcritical periodic external potential: $A=-0.0001$, $\Omega=\pi/4$, $\delta_{1,2,4,5}=0$, $\delta_3=\pi$; $\nu_3=0.5$, $\nu_{4,2}=\nu_3\pm0.01$, $\nu_{5,1}=\nu_3\pm0.02$;
polarization vectors $\theta_1=\frac{\pi}{3}$, $\theta_{k+1}=\theta_k-\frac{\pi}{8}$, $k=1,2,3,4$.}
\label{fig_r5b}
\end{figure}
\begin{figure}[h!]
\centerline{\includegraphics[width=0.9\textwidth]{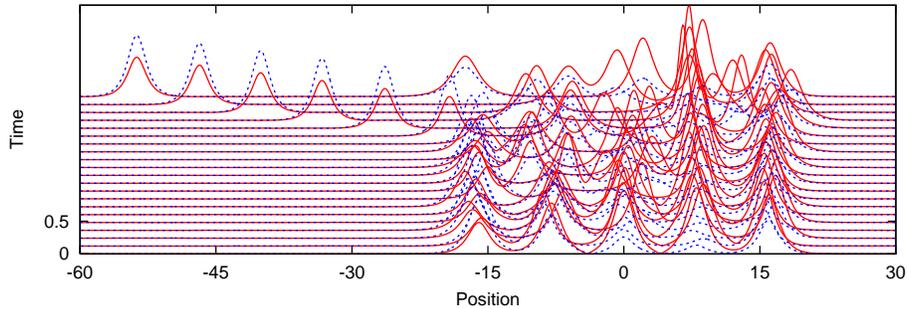}}
%\vspace{-0.1in}
\caption{Bound state asymptotic regime of 5-soliton envelopes. Perturbed two-component MS with linear gain $\gamma=10^{-4}$, cubic loss $\beta=-1.49\gamma$  and critical periodic external potential: $A=-0.0075$, $\Omega=\pi/4$, $\delta_{1,2,4,5}=0$, $\delta_3=\pi$; $\nu_3=0.5$, $\nu_{4,2}=\nu_3\pm0.01$, $\nu_{5,1}=\nu_3\pm0.02$; polarization angles $\theta_1=\frac{\pi}{3}$, $\theta_{i+1}=\theta_i-\frac{\pi}{8}$, $i=1,2,3,4$.}
\label{fig_r5c}
\end{figure}
\begin{figure}[h!]
\centerline{\includegraphics[width=0.45\textwidth]{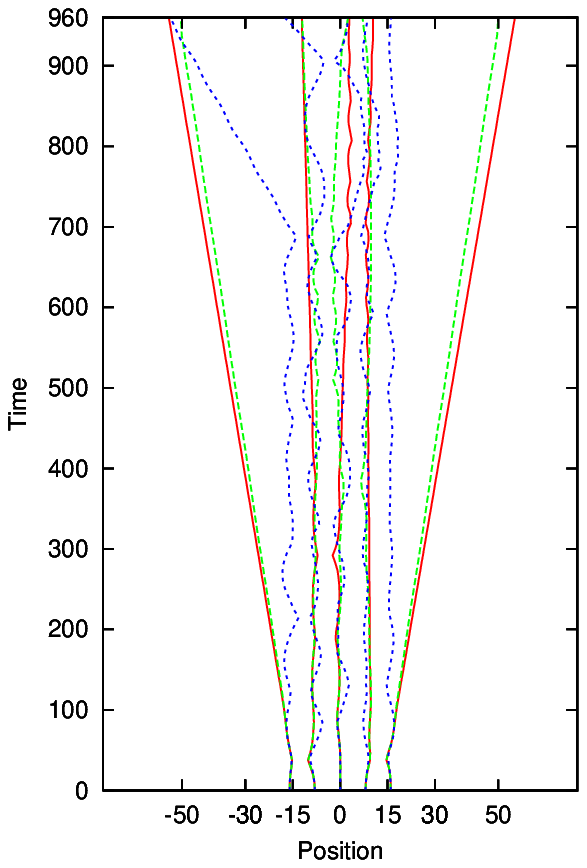}}
%\vspace{-0.1in}
\caption{5-soliton train -- Perturbed by a periodic potential two-component MS with linear gain $\gamma=10^{-4}$, cubic loss $\beta=-1.49\gamma$, $A=0$ (red), $A=-10^{-4}$ (green) and $A=-0.0075$ (blue), $\Omega=\pi/4$, $\delta_{1,2,4,5}=0$, $\delta_3=\pi$; $\nu_3=0.5$, $\nu_{4,2}=\nu_3\pm0.01$, $\nu_{5,1}=\nu_3\pm0.02$; polarization vectors $\theta_1=\frac{\pi}{3}$, $\theta_{i+1}=\theta_i-\frac{\pi}{8}$, $i=1,2,3,4$.}
\label{fig_r5d}
\end{figure}

\section*{Concluding Remarks and Future Activity}
A wide range of numerical experiments is conducted. All of them aim to report on the influence of linear and nonlinear
adiabatic perturbations (gain/loss + nonlinearity of 5-th degree) in the scalar nonlinear Schr\"odinger equation and the more generalManakov system. Besides above perturbations we consider their superposition with an external periodic potential.
We demonstrate that the gain/loss terms strongly affect the adiabatic approximation
and for generic choices of $\gamma$, $\beta$ and $\eta$ are not compatible with it. However we found special constraints
such as: (i) $\eta =0$, and $\beta \simeq -1.5 \gamma$; (ii) eq. (\ref{eq:ebgam}); after which
the gain/loss terms become compatible with adiabaticity. Then we can combine these gain/loss terms with a periodic potential
and show how one can switch over 5 asymptotically free solitons into a bound state, or into a 3-soliton bound state
and two out-going free solitons.

Other instruments to control the soliton interactions and the asymptotic behavior of the soliton trains
are based on the proper choice of the sets of soliton parameters and on the fact that the CTC is integrable, i.e.
according \cite{Moser} it possesses Lax representation $[L, M]=0$ (see also \cite{PRL,PRE,PLA}).
Given the initial soliton parameters we can evaluate the eigenvalues $\zeta_k = \zeta_{0k} + i \zeta_{1k}$ of $L$.
Since the real parts $\zeta_{0k}$  determine the asymptotic velocity of the $k$-th soliton, one can
determine the asymptotic regime of the soliton train. It will be a FAR of the solitons if all
$\zeta_{0k}$ are different; it will be a BSR if all $\zeta_{0k}$ are equal; it will be a MAR otherwise.

Quite similar facts hold true also for the Manakov system \cite{GDM,MCS,GT}.
Of course, often the GCTC is not integrable, so formally it does not posses Lax representation.
Nevertheless we may start from a soliton configuration that ensures, say a bound state regime and then
check whether the perturbation will alter it or not.

 Finally, we demonstrate that a large number of different types of perturbations can be effectively
treated as compatible with the adiabatic conditions. This, we believe, would enable large
 number of them to be effectively investigated.

\section*{Acknowledgements}
We are grateful to Prof. Ilia Iliev for useful discussions.

\appendix
\section{Typical integrals}

Here we list the typical integrals that appear in deriving the PCTC. First we list the integrals
\begin{equation}\label{eq:Jn0}\begin{split}
\int_{-\infty}^{\infty} \frac{dz}{\cosh^2(z)} = 2, \qquad \int_{-\infty}^{\infty} \frac{dz}{\cosh^4(z)} = \frac{4}{3},
\qquad \int_{-\infty}^{\infty} \frac{dz}{\cosh^6(z)} = \frac{16}{15}.
\end{split}\end{equation}
needed to derive $R_k[u]^{(0)}$.
It is possible to derive analytical expressions for all integrals $R_k[u]^{(1)}$, but these turn out to be
very involved. Below we are keeping only terms of the order of $\varepsilon_0$:
\begin{equation}\label{eq:Jn35}\begin{aligned}
\int_{-\infty}^{\infty} \frac{dz_k\; }{\cosh^3(z_k)\cosh (z_{k\pm 1})} &=  4 e^{-{|\Delta_{k,k\pm 1}|}}
+ \mathcal{O}(\varepsilon_0^{3/2}), \\
\int_{-\infty}^{\infty} \frac{dz_k\; }{\cosh^5(z_k)\cosh (z_{k\pm 1})} &=  \frac{8}{3} e^{-{|\Delta_{k,k\pm 1}|}}
+ \mathcal{O}(\varepsilon_0^{3/2}), \\
\int_{-\infty}^{\infty} \frac{dz_k\; \tanh (z_k)}{\cosh^3(z_k)\cosh (z_{k\pm 1})} &=  \mp \frac{4}{3} e^{-{|\Delta_{k,k\pm 1}|}}
+ \mathcal{O}(\varepsilon_0^{3/2}), \\
\int_{-\infty}^{\infty} \frac{dz_k\; \tanh (z_k) }{\cosh^5(z_k)\cosh (z_{k\pm 1})} &=  \mp \frac{8}{15} e^{-{|\Delta_{k,k\pm 1}|}}
+ \mathcal{O}(\varepsilon_0^{3/2}),
\end{aligned}\end{equation}
and
\begin{equation}\label{eq:Jn36}\begin{aligned}
 \int_{-\infty}^{\infty} \frac{dz_k\; z_k}{\cosh^3(z_k)\cosh (z_{k\pm 1})} &= \mp 2e^{-{|\Delta_{k,k\pm 1}|}}
+ \mathcal{O}(\varepsilon_0^{3/2}), \\
\int_{-\infty}^{\infty} \frac{dz_k\; z_k}{\cosh^5(z_k)\cosh (z_{k\pm 1})} &=  \mp \frac{2}{3} e^{-{|\Delta_{k,k\pm 1}|}}
+ \mathcal{O}(\varepsilon_0^{3/2}), \\
\int_{-\infty}^{\infty} \frac{dz_k\; (1 - z_k\tanh(z_k))}{\cosh^3(z_k)\cosh (z_{k\pm 1})} &= 2 e^{-{|\Delta_{k,k\pm 1}|}}
+ \mathcal{O}(\varepsilon_0^{3/2}), \\
\int_{-\infty}^{\infty} \frac{dz_k\; (1 - z_k\tanh(z_k))}{\cosh^5(z_k)\cosh (z_{k\pm 1})} &=  2 e^{-{|\Delta_{k,k\pm 1}|}}
+ \mathcal{O}(\varepsilon_0^{3/2}).
\end{aligned}\end{equation}

%where the coefficients $\mathcal{J}_{N,3,5}^\pm$, $\mathcal{J}_{M,3,5}^\pm$,
%$\mathcal{J}_{\Xi,3,5}^\pm$ and $\mathcal{J}_{D,3,5}^\pm$ are
%listed in the table below:
%\clearpage

\end{document}